\documentclass[10pt]{article}

\usepackage{amsmath,amssymb,amsthm}

\newcommand{\vtri}{\boldsymbol{\vartriangle}}

\begin{document}

\title{\bf  
An origin of spins of fields 
}

\author{Tsunehiro Kobayashi\footnote{E-mail: 
kobayash@a.tsukuba-tech.ac.jp} \\
{\footnotesize\it Research Center on Higher Education 
for the Hearing and Visually Impaired,}\\
{\footnotesize\it Tsukuba College of Technology}\\
{\footnotesize\it Ibaraki 305-0005, Japan}}

\date{}

\maketitle

\begin{abstract}

Spins of fields are investigated 
in terms of the zero-energy eigenstates of 2-dimensional 
Schr$\ddot {\rm o}$dinger equations with central potentials 
$V_a(\rho)=-a^2g_a\rho^{2(a-1)}$ ($a\not=0$, $g_a>0$ and  
$\rho=\sqrt{x^2+y^2}$). 
We see that for $a=N/2$ ($N=$positive odd integers) 
one half spin states can naturally be understood 
as states with the angular momentum $l=1$ in the $\zeta_a$ 
plane which is 
obtained by mapping the $xy$ plane in terms of 
conformal transformations $\zeta_a=z^a$ 
with $z=x+iy$. 
It is shown that the scalar and the 1/2-spin fields can 
obtain masses. 
Vortex structures and a supersymmetry for the zero-energy states 
are also pointed out. 

\end{abstract}

\thispagestyle{empty}

\setcounter{page}{0}

\pagebreak

\hfil\break
{\bf 1. Introduction}

\hfil\break
Particles generally have definite spins. 
It is well-known that 
quantum mechanics allows the existence of 
half integer spins like $J=1/2,3/2, 5/2,\cdots$. 
It is, however, 
 also well-known 
that the angular momentum for 
Schr\"odinger equations with central potentials 
can take only integers in non-relativistic quantum mechanics. 
To understand 1/2 spin naturally we need to write the equation of motion 
in terms of Dirac's relativistic equation. 
In this letter we shall see the fact that Schr\"odinger equations with 
some kind of two-dimensional central potentials can have the eigenstates that 
represent 1/2-spin states. 

Let us start with 
2-dimensional Schr$\ddot {\rm o}$dinger equations with the central potentials 
$V_a(\rho)=-a^2g_a\rho^{2(a-1)}$ ($a\not=0$ and  
$\rho=\sqrt{x^2+y^2}$). 
It has been shown that the equations 
have zero-energy eigenstates which are infinitely degenerate~
\cite{sk-jp,k-1,ks-pr,k-ps}. 
That is to say,  
the Schr$\ddot {\rm o}$dinger equations 
for the zero-energy eigenvalue 
are written as
\begin{equation}
 [-{\hbar^2 \over 2m}\vtri(x,y) +V_a(\rho)]\ \psi(x,y) 
  = 0, 
  \label{0}
\end{equation}
where 
$
\vtri(x,y)=\partial^2/ \partial x^2+\partial^2 / \partial y^2,
$  
and they are transformed into 
the following equation 
in terms of the conformal transformations 
$\zeta_a=z^a$ with $z=x+iy$
~\cite{k-1,ks-pr}; 
\begin{equation}
[-{\hbar^2 \over 2m}{\boldsymbol{\triangle}}_a-g_a]\ \ \psi_0 (u_a,v_a)=0, 
 \label{S0}
\end{equation}
where
$ 
{\boldsymbol{\triangle}}_a=\partial^2/ \partial u_a^2+
\partial^2/ \partial v_a^2.
$  
Here the variables are defined by 
the relations $\zeta_a=u_a+iv_a$ and 
\begin{equation}
 u_a=\rho_a \cos (\varphi_a), \ \ \ 
 v_a=\rho_a \sin (\varphi_a), 
 \label{uv}
\end{equation} 
where $\rho_a=\rho^a$ and $\varphi_a=a\varphi$. 
We see that the zero-energy eigenstates 
for all the different numbers of $a$ 
are described by the same plane-wave 
solutions in the $\zeta_a$ space. 
Furthermore it is easily shown 
that the zero-energy states degenerate infinitely.
Let us consider the case for $a>0$ and $g_a>0$. 
Putting the function $f^\pm_n (u_a;v_a)e^{\pm ik_a u_a}$ with 
$k_a=\sqrt{2mg_a}/\hbar $ 
into (2), 
where $f^\pm_n (u_a;v_a)$ are polynomials of degree $n$ ($n=0,1,2,\cdots $), 
we obtain the equations for the polynomials 
\begin{equation}
[{\boldsymbol{\triangle}}_a \pm2ik_a{\partial \over \partial u_a}]
f^\pm_n (u_a;v_a)=0.
\label{2}
\end{equation} 
Note that from the above equations we can easily see the relation  
$(f^-_n (u_a;v_a))^*=f^+_n (u_a;v_a)$ for all $a$ and $n$. 
General forms of the polynomials have been obtained by using the solutions 
in the $a=2$ case 
(2-dimensional parabolic potential barrier (2D PPB))~\cite{k-1,ks-pr}. 
Since all the solutions have the factors $e^{\pm ik_a u_a}$ 
or $e^{\pm ik_a v_a}$, we see that 
the zero-energy states describe stationary flows~\cite{sk-jp,k-1,ks-pr} 
and, of course, have no time factor. 
The general eigenfunctions 
with zero-energy are written as arbitrary linear combinations of the 
eigenfunctions included in two infinite 
series of $\left\{ \psi_{0n}^{\pm (u)}(u_a;v_a)  \right\}$ 
for $n=0,1,2,\cdots$, where 
\begin{equation} 
\psi_{0n}^{\pm (u)}(u_a;v_a)=
f_n^\pm(u_a;v_a)e^{\pm ik_a u_a}. 
\label{psi}
\end{equation}
(For the details, see the sections II and III of Ref.~\cite{ks-pr}.)
It has been also pointed out that  
the motions of the $z$ direction perpendicular to the $xy$ plane 
can be introduced as 
free motions represented by $e^{\pm ik_z z}$. 
In this case 
the total energies $E_T$ of the states are given by 
$E_T=E_z$, 
where $E_z$ are the energies of the free motions in the $z$ direction. 
Note that the zero-energy eigenfunctions cannot be normalized 
as same as those 
in scattering processes~\cite{bohm}. 
Actually it has been shown that all the zero-energy states for $g_a>0$ 
are eigenstates 
in the conjugate spaces of Gel'fand triplets~\cite{sk-jp,k-1,ks-pr,sk1}. 
In the recent work~\cite{k-gauge} it has been pointed out that 
the freedom for the infinite 
degeneracy can be understood as a kind of interacting gauge fields. 
In this letter we shall study the infinite degeneracy in terms of 
the eigenstates with the definite 
angular momentum in the $\zeta_a$ plane. 

\hfil\break
{\bf 2. Boundary conditions}

\hfil\break
Let us start our discussion from a simple example for $a=1/2$, 
because the simplest case for $a=1$ is trivial such that 
$u_1=x$ and $v_1=y$. 
As shown in Eq.~\eqref{uv}, the relation between the angle $\varphi_{1/2}$ 
in the $uv$ plane and the angle $\varphi$ in the $xy$ palne 
is given by 
$\varphi_{1/2}=\varphi/2$. 
This relation means that the whole of the $uv$ plane 
must be represented by two sheets of 
the $xy$ plane just as same as a Riemann surface~\cite{ks-pr,k-gauge}. 
For instance, the flow for 
the wave function $e^{-ik_{1/2}u_{1/2}(0)}$ has a cut 
from 0 to $\infty$ on the $x$ axis as shown in Fig. 1~\cite{k-gauge}. 
Actually 
all the zero-energy eigenstates given in Eq.~\eqref{psi} 
are not rotationally symmetric.  
We easily see that the eigenfunctions are classified into two types 
satisfying the definite boundary conditions for the rotation 
in the $xy$ plane such that 
\begin{align} 
\psi(\varphi+2\pi)_b&=\psi(\varphi), \ \ \ \ 
{\rm for\ bosonic\ boundary\ condition},
\nonumber  \\ 
 \psi(\varphi+2\pi)_f&=-\psi(\varphi), \ \ \ 
                  {\rm for\ fermionic\ boundary\ condition},
  \label{xy-bc}
\end{align} 
where the eigenfunctions are written as  
 \begin{align} 
\psi(\varphi)_b&=\psi(\varphi)+\psi(\varphi+2\pi), \ \ \ \ 
\nonumber  \\ 
 \psi(\varphi)_f&=\psi(\varphi)-\psi(\varphi+2\pi). \ \ \ 
  \label{bf}
\end{align} 
This means that we cannot decompose the eigenfunctions 
$ \psi(\varphi)_f$ in terms of the 
eigenfunctions of the angular momentum operator 
$-i\hbar \partial/\partial \varphi$ 
in the $xy$ plane, which are determined by the boundary condition 
$\psi(\varphi+2\pi)=\psi(\varphi)$. 
In the $uv$ plane, however, 
from the relations 
$u(\varphi_{1/2}+2\pi)=u(\varphi_{1/2})$ and 
$v(\varphi_{1/2}+2\pi)=v(\varphi_{1/2})$   
we see that all the eigenfunctions~\eqref{psi} 
satisfy the boundary condition  
\begin{equation}
\psi(\varphi_{1/2}+2\pi)=\psi(\varphi_{1/2}). 
  \label{uv-bc}
\end{equation} 
By considering the fact that the $2\pi$ rotation 
in the $uv$ plane corresponds to the $4\pi$ rotation in the $xy$ plane 
because of the relation $\varphi_{1/2}=\varphi/2$,  
we can write the Riemann sheet structure of the $xy$ plane 
as shown in Fig. 1. 
\begin{figure}
   \begin{center}
    \begin{picture}(200,200)
     \thicklines
    
     \put(0,100){\line(1,0){100}}
     \put(100,101){\line(1,0){100}}
     \put(100,99){\line(1,0){100}}
     \put(100,0){\vector(0,1){200}}
     \put(105,88){$0$}
     \put(205,98){$x$}
     \put(98,205){$y$}
     \put(98.6,97.6){$\bullet$}
     \put(130,105){${\rm cut}$}
 
     \put(102,120){\line(1,0){88}}
     \put(200,120){\vector(-1,0){10}}
     \put(102,80){\vector(1,0){98}}
     \qbezier(102,120)(80,100)(102,80)
     
     
     \put(200,100){\vector(1,0){3}}
    \end{picture}
   \end{center}
   \caption[]{Flow of the state expressed by 
   $e^{-ik_{1/2}u_{1/2}}$.}
   \label{fig:1a}
  \end{figure}
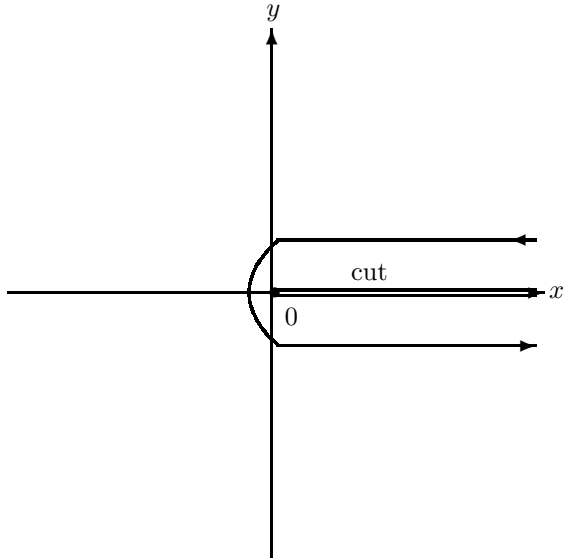
The fact that 
all the eigenfunctions satisfy the boundary condition 
of Eq.~\eqref{uv-bc} in the $uv$ plane indicates that 
the eigenfunctions can be written by those of the angular momentum 
with respect to the rotations for 
the angle $\varphi_{1/2}$ 
in the $uv$ plane. 

\hfil\break
{\bf 3. The angular momentum eigenstates in the $uv$ plane} 

\hfil\break
Let us here study the eigenstates for the angular momentum $\hat L_{uv}$ 
in the $uv$ plane, which are given by   
\begin{equation}
{\hat L}_{uv}=-i\hbar {\partial \over \partial \varphi_{1/2}}.
  \label{angular}
\end{equation} 
Thus the Schr\"{o}dinger equation (2) for the zero energy are written as 
\begin{equation} 
[-{1 \over q}{\partial \over \partial q}(q {\partial \over \partial q})
+ {1 \over q^2}({{\hat L}_{uv} \over \hbar})^2-g']\psi_l(q,\varphi_{1/2})=0,
\label{sch-psi}
\end{equation} 
where $q=\rho^{1/2}$ and $g'=2mg_{1/2}/\hbar^2$ are used. 
We can factorize the eigenfunctions as 
\begin{equation} 
\psi_l(q,\varphi_{1/2})=h_l(q)\phi_l(\varphi_{1/2}).
\label{h-phi}
\end{equation} 
It is trivial that 
the eigenfunctions for the angular momentum are given as usual 
\begin{equation}
\phi_l(\varphi_{1/2})=e^{ i l \varphi_{1/2}}/\sqrt{2\pi},
  \label{l}
\end{equation} 
which evidently satisfy the boundary condition~\eqref{uv-bc} 
in the $uv$ plane only for $l$=integers. 
Substituting Eq.~\eqref{h-phi} into Eq.~\eqref{sch-psi}, we obtain 
the equations for $h_l(q)$ as 
\begin{equation} 
[-{1 \over q}{\partial \over \partial q}(q {\partial \over \partial q})
+{l^2 \over q^2}-g']h_l(q)=0.
\label{hl}
\end{equation} 
From the condition that $h_l(q)$ have no singularity at 
$q=0$, 
we find out the solutions 
$$
h_l(q)=\sum_{n=0}^\infty c_{|l|+2n} q^{|l|+2n}, 
$$ 
where $c_n$ are determined by the following recurrence formula; 
\begin{equation}
c_{|l|+2n+2}=-{g' \over 4(n+1)(|l|+n+1)}\ c_{|l|+2n}=
{(-g' /4)^{n+1} \over (n+1)!(|l|+n+1)\cdots (|l|+1)}\ c_{|l|}. 
\label{c}
\end{equation} 
Note that the eigenfunctions are unique for each $|l|$ except $c_{|l|}$. 
Since the eigenfunctions in Gel'fand triplets are not normalizable in general, 
the constants $c_{|l|}$ should not be determined 
by the normalizations of $h_l(q)$~\cite{sk1}.  
Now we may conclude that the infinite degeneracy for the zero-energy 
eigenstates is expressed by the eigenstates of the angular momentum. 
It is strongly noticed that the infinite degeneracy can be represented as 
the infinite degeneracy of the states with arbitrary angular momenta 
in the $uv$ plane. 

\hfil\break
{\bf 4. Spins in the $xy$ plane}

\hfil\break
Let us study the meaning of 
the angular momentum eigenstates $\phi_l(\varphi_{1/2})$ 
in the $xy$ plane. 
Since the case for $l=0$ (scalar) is trivial, 
we study the case for $l=\pm1$. 
The eigenfunctions written by 
$\phi_{\pm 1}(\varphi)\equiv \phi_{\pm 1}(\varphi_{1/2})
=e^{\pm i\varphi/2}/\sqrt{2\pi}$ 
in the $xy$ plane 
are, of course, degenerate as the solutions 
of the original Schr\"odinger equation with the zero-energy eigenvalue. 
It is important that both of them satisfy the fermionic boundary condition 
$\phi_{\pm 1}(\varphi+2\pi)=-\phi_{\pm 1}(\varphi)$ in the $xy$ plane. 
Actually $\phi_{\pm 1}(\varphi)$ are the eigenfunctions of 
the angular momentum in the $xy$ plane, 
${\hat L}_{xy}=-i\hbar\partial/\partial \varphi$, of which 
eigenvalues are, respectively, $(\pm 1/2)\hbar $. 
We may say that $\phi_{\pm 1}$ 
stand for the spin-up and 
-down states for the spin 1/2 states, where the direction of 
the spin is the $z$ direction. 

Let us examine the solutions 
in relativistic motions.   
The above argument can be extended to the relativistic equations 
for massless particles 
in 4-dimensional (4D) Lorentz spaces. 
In the 4D spaces the separation of 
the two directions perpendicular to 
a non-zero 3-momentum ${\bf p}$ can be carried out 
in terms of the 4-momentums defined by $p^+=(|{\bf p}|,{\bf p})$ 
and $p^-=(-|{\bf p}|,{\bf p})$ such that 
$
\epsilon^{\mu \nu \lambda \sigma} p^+_\lambda p^-_\sigma,
$
where $\epsilon^{\mu \nu \lambda \sigma}$ is 
the totally antisymmetric tensor 
defined by $\epsilon_{0123}=1$. 
It is trivial to write $\rho=x^2+y^2$ and 
$
{\partial^2 \over \partial x^2} + {\partial^2 \over \partial y^2} 
$ 
in the covariant expressions in terms of these tensors. 
The Klein-Gordon equation for massless fields with the momentum of 
the $z$ direction can be written as 
\begin{equation} 
[{1 \over c^2}{\partial^2 \over \partial t^2} 
- \vtri(x,y,z)+V_a(\rho)]\psi_z(r)=0,
 \label{rel} 
\end{equation} 
where $\psi_z(r)=\psi(x,y)e^{i(p_zz-cp_ot)/\hbar}$ with $p_0^2=p_z^2$. 
Note that the dimension of $g_{1/2}$ in the 4D spaces 
is different from that in the 3D spaces. 
From the relation $p_0^2=p_z^2$ for the massless particles 
the derivatives with respect to $z$ and $t$ 
cancel out each other in Eq.~\eqref{rel}. 
Thus we see that $\psi(x,y)$ again satisfy the same equation as 
the zero-energy solutions in 2-dimensions, and 
$\psi(x,y)$ are obtained by replacing 
$k_a=\sqrt{2mg_a}/\hbar$ 
with $k_{1/2}=\sqrt{g_{1/2}}$ in the zero-energy eigenfunctions 
given by Eq.~\eqref{psi}. 
In the cases of the massless particles $\phi_{\pm1}$ are 
understood as the helicity eigenstates for the eigenvalues $h=(\pm1/2)\hbar $. 
Now we may say that so-called chiral fermions can quite naturally be 
understood in the present scheme. 
It is obvious that in the massless cases the states $\phi_l^\pm(\varphi_{1/2})$ 
are the helicity eigenstates for $l/2$ spin fields. 
Note that the eigenstates for $l$=even integers satisfy 
the bosonic boundary condition in the $xy$ plane, 
whereas those for $l$=odd integers 
do the fermionic one. 

Let us consider massive fields. 
In order to have a mass, the fields with spins $|l|/2$ 
must have $(|l|+1)$-fold degeneracy. 
Considering that these $(|l|+1)$-fold degenerate states must be represented 
by the same function for $h_l(q)$, we see that the scalar state for $l=0$ 
and the 1/2-spin state for $l=\pm 1$  are only the candidates for the 
massive fields. 
Since the scalar case is trivial, 
let us study the 1/2-spin fields. 
It is convenient to introduce the variables 
$$
\zeta=x+iy \ \ \ {\rm and}\  \ \ \zeta^* =x-iy.
$$ 
By using these variables we see that $q=(\zeta\zeta^*)^{1/4}$, 
$e^{i\phi}=(\zeta/\zeta^*)^{1/2}$, and 
the derivative are given by 
$$
\partial_\zeta=(\partial_x-i\partial_y)/2 \ \ \ {\rm and} 
\ \ \ \partial_{\zeta^*}=(\partial_x+i\partial_y)/2.
$$ 
We easily see that the operations of 
$\partial_\zeta$ and $\partial_{\zeta^*}$, respectively, 
transform the eigenstates for the angular momentum 
from $\phi_l$ to $\phi_{l\mp 1}$. 
Now the structure of the so-called Pauli spin matrices is trivially 
understood, and thus we can write the relativistic equation of motion for the 
1/2-spin fields for $p_z\not=0$ as 
\begin{equation} 
[\partial_0^2+(i{\boldsymbol{\sigma}}\cdot {\boldsymbol{\partial}})^2
+(mc/\hbar)^2+V_{1/2}(\rho)]\psi_{1/2}({\bf r},t)=0,
\label{massive}
\end{equation} 
where $\partial_0=(1/c)\partial_t$, and 
the transposed vector of $\psi_{1/2}$ are given by
\begin{equation} 
\psi_{1/2}^t({\bf r},t)=
     (\psi_{-1}(q,\varphi) ,\   \psi_1(q,\varphi ))
e^{i(p_z z-cp_0 t)/\hbar}
\label{mass-psi}
\end{equation} 
with $p_0^2=p_z^2+m^2 c^2$. 
Note that for the case of $a=1/2$ we have no simple form corresponding to 
Dirac's relativistic equations for massive 1/2-spin fields. 
The exception is the case of 2D PPB for the massless fields. 
We can write the equation of motion as 
\begin{equation} 
i\sigma^\mu (\partial_\mu -i A_\mu) \psi_{2}({\bf r},t)=0,
\
\label{mass-ppb}
\end{equation} 
where $\sigma^0$ is taken as the unite matrix, and 
\begin{equation}
A_\mu=c_0\epsilon^{\mu\nu\lambda\sigma} r_\mu p^+_\lambda p^-_\sigma 
\end{equation} 
with $c_0=$constant. 
By operating $-i\sigma^\mu (\partial_\mu +i A_\mu)$ to Eq.~\eqref{mass-ppb} 
we have a potential that is proportional to $\rho^2$. 
We see that only in the cases of the PPB 
the equation of motion for the massless fields 
can be written as the gauge interaction. 
It may be interesting to investigate the relations between the 2D structures 
composed of the zero-energy eigenstates for $a=2$ and 
the membrane in string theory.

\hfil\break
{\bf 5. Spins for $a=N/2$}

\hfil\break
We can extend the above discussions to the cases for $a=N/2$ 
with $N$= positive odd integers. 
The two boundary conditions 
for the angle $\varphi$ in the $xy$ plane given by Eq.~\eqref{xy-bc}
again appear and they are integrated into 
the same boundary condition for $\varphi_{a}$ as that 
given by Eq.~\eqref{uv-bc} 
in the $u_av_a$ plane. 
In these cases the Riemann surface composed 
of two sheets of the $xy$ plane correspond to 
$N$ sheets of the $u_av_a$ plane, that is to say, 
the $4\pi$ rotation  in the $xy$ plane corresponds 
to the $2N\pi$ rotation in the $u_av_a$ plane. 
We should notice that  
the boundary condition~\eqref{uv-bc} is defined by the $2\pi$ rotation 
in the $u_av_a$ plane. 
In order to change the $2N\pi$ rotation into the $2\pi$ rotation, 
we have to consider the rotation 
in the $u_av_a$ plane for $a=N/2$ by using 
the angle $\varphi_a/N$ instead of $\varphi_a$. 
Thus the angular-momentum eigenstates should be written by 
$e^{i l \varphi_a/N}=e^{il\varphi/2}$ that are same 
for all of $N$ in the $xy$ plane. 
Let us examine this situation by writing the current corresponding 
to the lowest eigenstate of Eq.~\eqref{psi} in the $xy$ plane, which is 
expressed by the corner flow with the angle $\pi/a$~\cite{sk-jp,ks-pr}. 
An example for $a=3/2$ is shown in Fig. 2~\cite{k-gauge}, 
where three corner flows 
with the angle $2\pi/3$ contained in a $xy$ plane are figured. 
In general for $a=N/2$ 
the $N$ number of corner flows are contained 
in a $xy$ plane. 
\begin{figure}
   \begin{center}
    \begin{picture}(200,200)
     \thicklines
    
     \put(0,100){\vector(1,0){100}}
     \put(100,101){\line(1,0){100}}
     \put(100,99){\line(1,0){100}}
     \put(100,0){\vector(0,1){200}}
     \put(105,88){$0$}
     \put(205,98){$x$}
     \put(98,205){$y$}
     \put(98,98){$\bullet$}
     \put(120,105){${\rm cut}$}
 
     \put(190,110){\vector(1,0){10}}
     \qbezier(55,190)(105,115)(190,110)
     
     \put(64,102.5){\vector(0,-1){5}}
     \qbezier(40,180)(85,100)(45,20)
     
     \put(200,90){\vector(-1,0){10}}
     \qbezier(65,10)(105,85)(190,90)
     
    \end{picture}
   \end{center}
   \caption[]{Three corner flows for $a=3/2$ in a $xy$ plane.}
   \label{fig:2a}
  \end{figure}
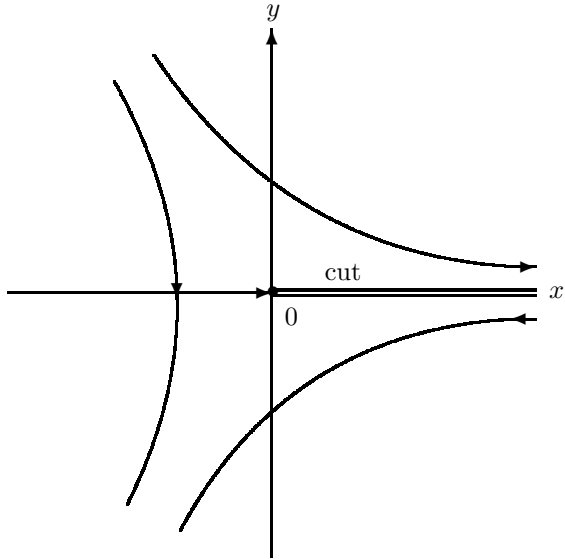
We see that  
one incoming (or outgoing) current composed of the two corner flows 
is contained in 
one sheet of the $N$ sheets of the $u_av_a$ plane. 
The diagram composed of the $2N$ corner flows 
in one Riemann surface of the $xy$ representation (in the double sheets) 
is invariant with respect to the $4\pi/N$ rotation that exactly corresponds 
to the $2\pi$ rotation in the $u_av_a$ plane. 
Note that in the consideration of the invariance the coincidence 
of the directions of the flows must be taken into account. 
For instance the $2\pi/N$ rotation change the directions of the flows. 

\hfil\break
We have seen that the half-integer spins originate from the 2-dimensional 
structure described by the zero-energy states of the potentials $V_a(\rho)$. 
It should strongly be noticed that, since the zero-energy states have no 
time development at all, the spin structures are absolutely stable in 
free motions. 
The change of the structure will occur in the interaction with gauge fields 
induced by the zero-energy states~\cite{k-gauge}. 

\hfil\break
{\bf 6. Remarks} 

\hfil\break
Finally we shall comment on two interesting properties of the zero-energy 
solutions. 

\hfil\break
(1) Vortices of the fields with spins
\hfil\break
Let us investigate the vortex structure of the eigenfunctions 
$\psi_l(q,\varphi)$ given in Eq. (11) for $a=1/2$. 
It is known that the angular momentum eigenstates have a singularity 
in the velocity at the origin ($\rho=0$), and then a vortex appears 
at the origin~\cite{sk-jp,k-1,ks-pr}. 
Let us evaluate the circulation represented by 
the integration with respect to a closed circle $C$ 
encircling the singurlarity as 
\begin{equation}
\Gamma_l=\oint_C {\bf v}_l\cdot d{\bf s},
\end{equation} 
where the velocity is defined by 
${\bf v}_l={\cal R}e [\psi_l^*(-i\hbar{\boldsymbol{\partial}}\psi_l]/|\psi_l|^2$
~\cite{sk-jp}. 
We obtain 
\begin{equation}
\Gamma_l={\pi \hbar \over m}\ l 
\end{equation} 
in the non-relativistic case. 
It is interesting that the circulation is quantized 
in the unite of $\pi \hbar/m$, while it is quantized in the unite 
$2\pi \hbar/m$ for the eigenstates of the angular momentum 
for central potentials, which belong to Hilbert spaces. 
(For example, see Appendix A3 of Ref.~\cite{sk-jp}.) 
This difference is easily understood from the fact that 
we obtain the same quantization with the unite $2\pi \hbar/m$ 
for the circulation in the $u_av_a$ plane. 
That is to say, since the $2\pi$ rotation in the $u_av_a$ plane corresponds 
to the $4\pi$ rotation in the $xy$ plane, the circulation defined by 
the $2\pi$ rotation in the $xy$ plane is quantized in the half of 
the unite $2\pi \hbar/m$ obtained in the $u_av_a$ plane. 
Similar arguments can be carried out in the cases of $a>1/2$.

\hfil\break
(2) Supersymmetry of the zero-energy states 
\hfil\break
Finally we would like briefly comment on supersymmetry. 
We have seen that in the cases of $a=N/2$ bosonic states and fermionic 
states appear alternately in increasing $|l|$. 
This fact indicates that the total system composed of the 
zero-energy states have some kind of supersymmetry. 
We can understand the supersymmetric property 
of the zero-energy states 
as follows; 
\hfil\break
All the zero-energy eigenfunctions~\eqref{psi} can be written down by 
the two types corresponding to bosonic and fermionic states 
given in Eq.~\eqref{bf} such that 
\begin{align} 
\psi_{0n(b)}^\pm &=\psi_{0n}^\pm (\varphi)+\psi_{0n}^\pm(\varphi+2\pi),
 \ \ \ \ {\rm for\ bosonic\ boundary\ condition} 
\nonumber \\
\psi_{0n(f)}^\pm &=\psi_{0n}^\pm(\varphi)-\psi_{0n}^\pm(\varphi+2\pi),
 \ \ \ {\rm for\ fermionic\ boundary\ condition}. \nonumber 
  \label{xy-bc2}
\end{align}  
This fact means that every zero-energy eigenstates can be written by 
the sum of the bosonic and fermionic states as 
$$
\psi_{0n}^\pm(\varphi)={1 \over 2}[\psi_{0n(b)}^\pm
  +\psi_{0n(f)}^\pm].
  $$ 
In these sums the weights for the bosonic state and the fermionic 
one are the same value 1/2. 
Except the states having definite angular momenta like $\psi_l$ 
all stationary phenomena described by the zero-energy eigenstates~\eqref{psi} 
are expected to have a supersymmetry with respect to the exchange 
between the bosonic and fermionic states.

\pagebreak


\begin{thebibliography}{1}



 
 


\bibitem{sk-jp} T. Shimbori and T. Kobayashi, 
     \emph{J. Phys.}  {\bf A33} 7637 (2000).
 
\bibitem{k-1} T. Kobayashi, 
     \emph{Physica}  {\bf A303} 469 (2002).
 
\bibitem{ks-pr} T. Kobayashi and T. Shimbori, 
     \emph{Phys. Rev.}  {\bf A65} 042108 (2002). 

\bibitem{k-ps} T. Kobayashi, 
     \emph{Phys. Scripta} {\bf 70} 335 (2003). 



 \bibitem{bohm}
	  A. Bohm and M. Gadella, 
	  \emph{Dirac Kets, Gamow Vectors and Gel'fand Triplets} 
	  (Lecture Notes in Physics, Vol. 348, Springer, 1989). 
	  

  \bibitem{sk1}
  Shimbori~T and Kobayashi~T, 
      \emph{Nuovo Cim.} {\bf 115B} 325 (2000). 

\bibitem{k-gauge} T. Kobayashi, 
     Interacting gauge fields and the zero-energy eigenstates in 
     two dimensions, 
     e-print, het-th/0503051 (2005). 
     

\end{thebibliography}
\end {document}